\documentclass{elsart}

 \usepackage{graphicx}
\usepackage{amssymb,amsmath}
\usepackage{bm}% bold math
\usepackage{dcolumn}% Align table columns on decimal point
\usepackage[dvips]{color}
\usepackage{ulem}

\begin{document}

\begin{frontmatter}



\title{Surface tension of membranes depending on the boundary shape}


\author{Hiroshi Koibuchi}
\ead{koibuchih@gmail.com}

\address{Department of General Engineering, National Institute of Technology (KOSEN), Sendai College, 
48 Nodayama, Medeshima-Shiote, Natori-shi, Miyagi 981-1239, Japan}

\begin{abstract}
In this paper, we study the boundary effect on the surface (or frame) tension of elastic membrane surface models. The frame tension generally depends only on the projected area of the boundary over which the surface spans. However, from a spin model analogy, the frame tension is expected to be dependent also on the boundary shape at the continuous transition point. We confirm this expectation using the following fixed-connectivity and tethered surface models: the surface model of Helfrich and Polyakov and a surface model with deficit angle term. We also discuss the reason why this expectation is worthwhile to study. 
\end{abstract}

\end{frontmatter}

% main text
%\section{}
%\label{}

%----------------------------------------------------------
\section{Introduction\label{intro}}
%----------------------------------------------------------
The elastic surface model of Helfrich and Polyakov is defined on triangulated lattices \cite{HELFRICH-1973,POLYAKOV-NPB1986,KANTOR-NELSON-PRA1987-1,KANTOR-NELSON-PRA1987}, and many theoretical and numerical studies have been conducted based on this model \cite{Bowick-PREP2001,WIESE-PTCP19-2000,NELSON-SMMS2004,GOMPPER-KROLL-SMMS2004,Essa-Kow-Mouh-PRE2014,KD-PRE2002,Nishiyama-PRE2010-2}. This surface model is a two-dimensional natural extension of linear chains for one-dimensional object polymers \cite{Doi-Edwards-1986}. In the triangulated lattice models for membranes, the response to the applied mechanical force is calculated as the surface (or frame) tension by fixing the boundary vertices from the scale invariant property of the partition function \cite{WHEATER-JP1994}. In this paper, we show that the frame tension of fixed-connectivity (or tethered) models  depends on the shape of boundary at a continuous transition point of surface fluctuations or crumpling transition. (See Section \ref{tethered-fluid} for more detailed information on "tethered" and "fluid" surface models defined on triangulated surfaces.)

The frame tension $\sigma$ is expected to depend on the shape of the surface boundary; in other words, the mechanical property of materials depends on their shape. Indeed, $\sigma$ is strongly influenced by the thermal fluctuations and hence by phase transitions \cite{Kadanov-etal--RMP1967}. Since the phase transitions are generally influenced by the boundary conditions imposed on the dynamical variable, the boundary shape influences the frame tension of the membranes. The boundary shape difference is expected to cause a nontrivial influence on the phase transition, to which all of the dynamical surface variables contribute, where the surface position ${\bf r}(\in {\bf R}^3)$ is the dynamical variable in the case of fixed-connectivity (or tethered) surface model. Therefore, it is very interesting to study whether $\sigma$ depends on the boundary shape \cite{Koibuchi-PLA2016}. 

On the tethered surfaces, not only bending resistance but also shear resistance is expected to appear in the surface deformation. Hence, the presence of this shear resistance influences  the surface fluctuations. For this reason, if the fluctuation pattern is changed by the boundary shape, this change in the fluctuations may cause a non-trivial difference in the frame tension if the surface is relatively smooth. This is in sharp contrast to the case of fluid membranes, where the shape dependence of $\sigma$ is not expected in general \cite{Cai-Lub-PNelson-JFrance1994,Dobreiner-et-al-PRL2003,PDPJB-EPJE2004,Fournier-PRL2008,Fournier-PRL2004,David-Leibler-JPF1991,Foty-etal-PRL1994,Foty-etal-Devlop1996}, because no shear resistance is expected due to the free diffusion of vertices on fluid surfaces. 
In this paper, the frame tension dependency on the boundary shape is verified \cite{Koibuchi-PLA2016}. To see this boundary influence on the frame tension $\sigma$, we assume two different shapes for the boundary in the simulations, where no anisotropy is assumed in the models of this paper \cite{Noguchi-PRE2011}. 

Here we emphasize the reason why it is interesting to study the dependence of the frame tension on the boundary shape. We should note that the variable, the position ${\bf r}$ of an arbitrary lattice point or vertex, becomes dependent on the boundary shape only when the system undergoes continuous transition. Indeed, using a spin model analogy, the variables on the surface boundary influence all of the lattice variables at the continuous transition point \cite{PL-PRL1985,GDLP-PRL1988,DG-EPL1988}, where the correlation length  (or persistence length \cite{DeGennes-JPC1982}) $\xi$ is expected to be divergent at the bending rigidity $\kappa_c$:  

\begin{eqnarray}
\label{divergent-corr-length}
\xi\to\infty\quad  (\kappa\to \kappa_c \; :\; {\rm surface\; model}),
\end{eqnarray}
which corresponds to the expectation in spin models such as $\xi\!\to\!\infty\; (T\!\to\! T_c : {\rm spin\; model} )$ at the critical temperature  $T_c$ \cite{Parisi-SFT1988}. 
In the case of surface models, the persistence length of the surface is connected to the surface normal vector ${\bf n}$ such that $\langle {\bf n}(0)\!\cdot\!{\bf n}(r)\rangle$, which is expected to behave $\langle {\bf n}(0)\!\cdot\!{\bf n}(r)\rangle\!\sim\! r^{-\eta}$ with a critical exponent $\eta$ at the continuous transition point $\kappa_c$. This power low decay is numerically confirmed by Monte Carlo (MC) simulations on free boundary surfaces \cite{Bowick-1996JPF}, and this is not an exponential decay, and hence the persistence length $\xi$ becomes infinite for $r\!\to\!\infty$. Moreover, the surface normal  ${\bf n}$ is not the dynamical variable but is only connected to a second order differential of the variable ${\bf r}$ \cite{FDavid-SMMS2004}. As a consequence, the fluctuation of ${\bf n}$ is not always identical to the fluctuation of the surface position ${\bf r}$. In fact,  ${\bf n}$ remains unchanged whenever the variable ${\bf r}$ fluctuates only into the in-plane directions, and  ${\bf n}$ varies only when the variable ${\bf r}$ fluctuates into the out-of-plane direction. This is in sharp contrast to the case of spin models, where the spin variable itself is the dynamical variable.  

Therefore, it is non-trivial and worth while to study whether the boundary variables influence all of the surface variables $\{{\bf r}\}$ at the continuous transition point of the surface models. In other words, if the persistence  length is divergent on a surface with boundary, then the variables ${\bf r}$ fixed at the boundary influence $\{{\bf r}\}$ on the whole surface.  Therefore, it is natural to consider that this influence of boundary is reflected in the mechanical property such as $\sigma$.  The problem is whether $\sigma$ depends on the boundary shape or not. Moreover, such a non-trivial dependence of $\sigma$ on the boundary shape is expected on the tethered surfaces because of non-zero shear resistance as we mentioned above \cite{Cai-Lub-PNelson-JFrance1994,Dobreiner-et-al-PRL2003,PDPJB-EPJE2004,Fournier-PRL2008,Fournier-PRL2004,David-Leibler-JPF1991,Foty-etal-PRL1994,Foty-etal-Devlop1996}.   

If the frame tension $\sigma$ is independent of the boundary shape under the above-mentioned property in Eq. (\ref{divergent-corr-length}), the tethered surface model defined on fixed-connectivity lattices is expected to be close to the fluid surface model on dynamically triangulated lattices at the continuous transition point. In contrast, if $\sigma$ depends on the boundary shape, the tethered and fluid surface models defined on triangulated lattices are different as expected, and moreover the continuous transition of the tethered surface model shares the same property as that of the spin models.
 
We should note that the shape dependent surface tension in this paper is different from anisotropic surface tension \cite{Noguchi-PRE2011,Steven-etal-JID1989,Arroyave-etal-ProEng2015}. In fact,  the anisotropic surface tension is simply direction dependent and not always shape dependent, and conversely it is clear that the shape dependent surface tension is not always anisotropic, although materials with anisotropic surface tension appear to have a shape-dependent surface tension \cite{Steven-etal-JID1989,Arroyave-etal-ProEng2015}. 

Note also that the calculation technique of frame tension can also be applied to three-dimensional polymeric materials such as liquid crystal elastomers \cite{Koibuchi-POL2017}. Therefore, this technique can be used to study how the macroscopic mechanical property is connected to microscopic processes such as thermal fluctuations or nematic transition of the constituent molecules \cite{Mblanga-etal-PRE2010,Wei-etal-PRE2011,Corbett-Warner-SensActu2009,Yusuf-etal-PRE2005,Na-etal-PRE2011}. 
%----------------------------------------------------------
\section{Models\label{models}}
%----------------------------------------------------------
%----------------------------------------------------------
\subsection{Triangulated lattices\label{lattice}}
%----------------------------------------------------------
%%%%%%%%%%%%%%%%%%%%%%%%%%%%%%%%%%%%%%%%%%%%%%%%%%%%%%%%%%%%%%%%%%%%
\begin{figure}[t]
\begin{center}
\includegraphics[width=8.5cm]{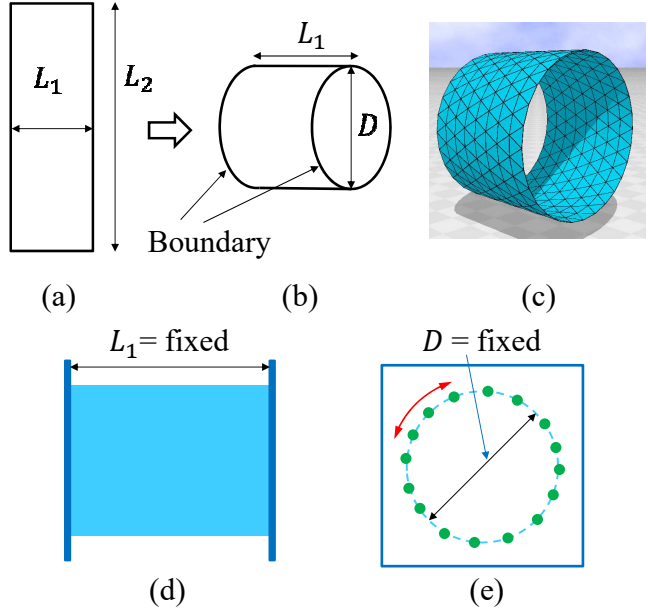}
 \caption{(a) A rectangle of size $L_1\sqrt{3}/2\times L_2$, (b) the cylinder of height $H\!=\!L_1\sqrt{3}/2$ and diameter $D\!=\!L_2/\pi$,  (c) a triangulated lattice of $(L_1,L_2)\!=\!(11,29)$, which approximately makes $D/L_1\!=\!1$, (d) a front view of the cylinder with fixed boundary, and (e) the vertices (\textcolor{green}{$\bullet$}) on a boundary are allowed to move along the circumferential direction. The square plates in (d) and (e) are drawn to emphasize the existence of the fixed boundaries. The edge length $a$ of the regular triangle in (c) is assumed as $a\!=\!1$.}
 \label{fig-1}
\end{center}
\end{figure}
%%%%%%%%%%%%%%%%%%%%%%%%%%%%%%%%%%%%%%%%%%%%%%%%%%%%%%%%%%%%%%%%%%%%
We use triangulated cylinders of two different ratios $D/L_1$ of diameter $D$ and $L_1$ such as $D/L_1\!=\!1$ and  $D/L_1\!=\!4$. Figure \ref{fig-1} shows how to construct the lattices, where the symbols $L_1$ and $L_2$ in Fig. \ref{fig-1}(a) denote the total number of lattice points along horizontal and vertical directions. Let $a$ be the lattice spacing, which is the edge length of the regular triangle; then, we have the diameter $Da(=\!L_2a/\pi)$ and the height $Ha(=\!L_1\sqrt{3}/2)$ in Fig. \ref{fig-1}(b). A cylinder of size $(L_1,L_2)\!=\!(9,28)$ is shown in Fig. \ref{fig-1}(c), where the ratio $D/L_1$ is approximately given by $D/L_1\!=\!1$. We should note that the real length for $L_1$ is given by $aL_1\sqrt{3}/2$ if $a$ is included, whereas $aD$ is the real diameter. In the following, the lattice spacing $a$ is fixed to $a\!=\!1$ for simplicity.

The boundary condition assumed for the calculation of frame tension is as follows: The distance $L_1$ between the two boundaries is fixed, and the diameter $D$ of the boundary is also fixed (Figs. \ref{fig-1}(d),(e)). Because of these conditions the vertices of the boundaries are prohibited from moving freely, however the vertices are not completely fixed and allowed to move along the circumferential direction (Fig. \ref{fig-1}(e)). We should note that this constraint is imposed only on the vertices on the boundaries, of which the total number is given by $2L_2$ as described above.

From the boundary condition imposed on the cylinders we expect that the two different ratios $D/L_1\!=\!1$ and  $D/L_1\!=\!4$ play a role of two different shapes of boundary frame of plates such that $L_2/L_1\!=\!\pi $ and $L_2/L_1\!=\!4\pi$, respectively.

Lattices used in the simulations are listed in Table \ref{table-1}. We use five different-sized lattices in each of two different shapes characterized by $D/L_1\!=\!1$ and  $D/L_1\!=\!4$. $N_{\rm E}(=\!N\!-\!2L_2)$ is the total number of vertices except for the boundary vertices. 

%++++++++++++++++++++++++++++++++++
\begin{table}[hbt]
\caption{Data of the lattices: the total number of vertices $N(=\!L_1L_2)$; $L_1$, $L_2$, $N_{\rm E}(=\!N\!-\!2L_2)$, $D(=\!L_2/\pi)$ and $H(=\!L_1\sqrt{3}/2)$. The lattices are grouped into two types, $D/L_1\!=\!1$ and  $D/L_1\!=\!4$.}
\label{table-1}
\begin{center}
 \begin{tabular}{cccccccc}
 \hline
$N$ & $L_1$ & $L_2$ & $N_{\rm E}$ & $D$ & $H$ & $D/L_1$ & $L_2/N$  \\
 \hline
   60604 & 139 & 436 & 59732 & 138.8 & 120.4 & 1 & 0.719  \\
   32017 & 101 & 317 & 31383 & 100.9 &  87.5 & 1 & 0.990  \\
   16790 &  73 & 230 & 16330 &  73.2 &  63.2 & 1 & 1.37   \\
    8798 &  53 & 166 &  8466 &  52.8 &  45.9 & 1 & 1.89   \\
    4758 &  39 & 122 &  4680 &  38.8 &  33.8 & 1 & 2.56   \\
 \hline
   59823 & 69  & 867 & 58089 & 276.0 &  59.8 & 4 & 1.45   \\
   30184 & 49  & 616 & 28952 & 196.1 &  42.4 & 4 & 2.04   \\
   15400 & 35  & 440 & 14520 & 140.1 &  30.3 & 4 & 2.86   \\
    9153 & 27  & 339 &  8457 & 107.9 &  23.4 & 4 & 3.70   \\
    4541 & 19  & 239 &  4063 &  76.1 &  16.5 & 4 & 5.26   \\
 \hline
\end{tabular}
\end{center}
\end{table}
%++++++++++++++++++++++++++++++++++

%----------------------------------------------------------

\subsection{Introduction to triangulated surface models\label{tethered-fluid}}
%----------------------------------------------------------
In this subsection, we briefly make general and introductory remarks on the surface models defined on triangulated lattices. First, we intuitively  explain the terminologies "tethered", "fluid" and "self-avoiding", "non-selfavoiding" surfaces, the first two of which are already used in the Introduction.  These words are used only for membranes. In this paper, we study tethered and non-self-avoiding surface models, which will be introduced in the following subsections.

The triangulated surface models are divided into two groups; the tethered and fluid models, where the vertices are considered to be lipids or group of lipids \cite{HELFRICH-1973}. The surface model in which the vertices are always connected by the edges or bonds is called "tethered" model (Fig. \ref{fig-1}(c)), while the model in which the vertices can diffuse over the surface is called "fluid" model. The diffusion of vertices in the fluid model is realized on dynamically triangulated lattices, where the bonds are flipped as one of the MC steps \cite{AMBJORN-NPB1993,Ho-Baum-EPL1990}. As we immediately understand, this bond flip MC process changes the lattice structure dynamically. In this sense the lattice structure itself in the fluid model is considered as a dynamical variable.  This dynamical variable has been considered a discrete analogue of the metric $g_{ab}$ degree of freedom, which is a function to be integrated out in the partition function in the continuous surface model \cite{POLYAKOV-NPB1986}.

Therefore, the "fluid" surface does not always share the same property with the standard fluids. The standard fluids are incompressible, while the fluid (and also tethered) surface is compressible. However, because of the so-called scale invariant property mentioned below, the mean surface area remains constant even when the surface has no fixed boundary; the property that the mean area remains constant is independent of whether the surface has the boundary or not. In this sense the surface models on triangulated lattices are considered to be "incompressible" as a two-dimensional material. Moreover, this property for the constant area can be seen even when the surface is folded.

The fluid surface model has no resistance for shear deformation like the standard fluids because of the free diffusion of vertices. To the contrary, the vertices always move only locally in the tethered model, and hence the shear resistance is naturally expected.

The meaning of "fluid" is limited to triangulated surface models and is slightly different from that of molecular dynamics (MD) simulation models. In MD simulations, there is no tethered model and all models are fluid, because the lipid molecules are allowed to diffuse over the surface in MD  \cite{Noguchi-PRE2011}. In contrast, the vertices, which are considered to be lipids or group of lipids as mentioned above, are connected by the edges as shown in Fig. \ref{fig-1}(c) even in the fluid models. One advantageous side of  triangulated lattice models is that we can use geometric or rigorous mathematical notions such as curvature etc. to define the models for membranes.

Another grouping of the triangulated surface models is; "self-avoiding" (SA) and  "non-selfavoiding" (non-SA) models. A surface model is called SA if the surface is prohibited from self-intersecting \cite{KANTOR-NELSON-PRA1987}. However, the SA potential is non-local, and for this reason, MC simulations for the SA models are very time consuming, and the lattice size is still limited in their MC studies. To the contrary, a surface model which is not SA is called non-SA or self-intersecting.  However, if the non-SA surface is sufficiently smooth, the surface will not be self-intersected. For this reason, to make the surface rigid, curvature energies such as extrinsic curvature or intrinsic curvature are introduced with the stiffness or bending-rigidity constant, which will be described in the following subsections.

 Non-SA surface models with curvature energy are expected to be non-folding or almost self-avoiding if $\kappa>\kappa_c$, where $\kappa_c$ is the crumpling transition point \cite{Essa-Kow-Mouh-PRE2014,KD-PRE2002}. 
In contrast, the vertices of the non-SA models can occupy the same region in ${\bf R}^3$ for sufficiently small bending rigidity ($\kappa<\kappa_c$) if the surface has no fixed boundary. In the case of surfaces with the fixed boundary like the models in this paper, we expect that this unphysical situation does not appear even at the transition point  $\kappa\!=\!\kappa_c$.

The order of the crumpling transition of tethered and non-SA surface model without fixed boundary is still controversial \cite{Cuern-etal-PRE2016}. If the transition of this surface model is of first order, the property in Eq. (\ref{divergent-corr-length}) is not expected. However, the transition is always expected to become weak and continuous in the presence of fixed boundary even when it is a strong such as first-order transition on free boundary surfaces. This is the reason why we study tethered and non-SA surface models with fixed boundary.

We should emphasize that the vertices diffuse freely over the surface in the fluid models as described above \cite{AMBJORN-NPB1993,Ho-Baum-EPL1990}. As a consequence the frame tension may be independent of the boundary-shape. On the contrary, the vertex fluctuates only locally in the case of tethered model, and therefore, we expect that the frame tension is boundary-shape dependent in the models of this paper. 

%----------------------------------------------------------
\subsection{Canonical surface model\label{Cano}}
%----------------------------------------------------------
The discrete surface model is obtained by a discretization of the Helfrich and Polyakov continuous model for membranes and by the assumption that the metric function $g_{ab}$ of the surface is given by the Euclidean metric $\delta_{ab}$. For this case of $g_{ab}\!=\!\delta_{ab}$ we call the model the {\it canonical} model; the Hamiltonian $S$ is given by a linear combination of the Gaussian bond potential $S_1$ and the bending energy $S_2$ such that \cite{KANTOR-NELSON-PRA1987-1}
\begin{eqnarray}
\label{Hamiltonian-can} 
&& S({\bf r})=\lambda S_1 + \kappa S_2, \quad(\lambda=1),  \nonumber \\
&& S_1=\sum_{ij} \left( {\bf r}_i-{\bf r}_j\right)^2, \quad S_2=\sum_{ij} (1-{\bf n}_i \cdot {\bf n}_j), \qquad ({\rm canonical}).
\end{eqnarray} 
The symbol ${\bf r}$ in $S$ denotes the vertex position such as ${\bf r}=\{{\bf r}_1,{\bf r}_2,\cdots,{\bf r}_N\}$, where $N$ is the total number of vertices including $2L_2$ vertices on the boundary. The surface tension parameter $\lambda$ is fixed to $\lambda\!=\!1$. We should note that the physical unit of $\lambda$ is ${\rm [k_BT/m^2]\!=\![N/m]}$, because $S_1$ has the unit of $a^2{\rm  [m^2]}$ with the lattice spacing $a$, which is an adjustable parameter  \cite{Creutz-txt}, and $k_B$ is the Boltzmann constant and $T$ is the temperature. Here in this paper, $a$ is fixed to $a\!=\!1$ as mentioned above, because the calculated physical quantity is not compared with experimental data. The sum $\sum_{ij}$ in $S_1$ is over all nearest neighbor vertices $i$ and $j$, connected by a triangle edge. The coefficient $\kappa[{\rm 1/k_BT}]$ is the bending rigidity. The symbol ${\bf n}_i$ in $S_2$ denotes a unit normal vector of the triangle $i$, and $ij$ in the sum $\sum_{ij}$ denotes the two nearest neighbor triangles sharing the common bond $ij$, and the boundary bonds $ij$ are not included in this $\sum_{ij}$. 

Note that the Gaussian curvature term $S_{\rm GC}$ is included in the original model of Helfrich in Ref. \cite{HELFRICH-1973}, and   $S_{\rm GC}$ is expected to play an important role in the case of closed surfaces of which the surface topology changes. This term $S_{\rm GC}$ is eliminated from the Hamiltonian for simplicity, because the surface assumed this paper is not closed and the topology is not changed  or holes are not assumed on the surface. 

The partition function $Z$ is given by 
\begin{eqnarray}
\label{part-func}
&&Z(A_p) = \int \left(\prod _{i=1}^{N_{\rm E}} d {\bf r}_i\right) \left(\prod _{i=1}^{2L_2} d {\bf r}_i\right)^\prime \exp\left[-\beta S({\bf r})\right],  \nonumber \\ &&N_{\rm E}=N-2L_2,
\quad (\beta=1),
\end{eqnarray}
where $\int\left(\prod _{i=1}^{N_{\rm E}} d {\bf r}_i\right)$ denotes the three-dimensional multiple integrations for the vertices except the boundary vertices, and $\int\left(\prod _{i=1}^{2L_2} d {\bf r}_i\right)^\prime$ denotes the one-dimensional multiple integrations for the boundary vertices, which are allowed to move on the boundary circle as described in Section \ref{lattice} so that the boundary shape remains unchanged. $A_p$ in $Z(A_p)$ is the projected area of the cylinder and is given by $A_p\!=\!\pi D L_1\sqrt{3}/2\!=\!L_1L_2\sqrt{3}/2$, which remains unchanged even when the surface area changes by thermal fluctuations. The inverse temperature $\beta(=\!1/k_BT)$ is fixed to be $\beta\!=\!1$. 

We should note that the unit of $\kappa$ is given by $[1/k_BT]$ as mentioned above. This is a consequence of the convention $\beta\!=\!1$, and from this we consider that $\kappa \to \infty$ corresponds to $T\to 0$ (or $\kappa \to 0$ corresponds to $T\to \infty$). This implies that the surface becomes rigid (soft) for sufficiently low (high) temperature. 
At the same time the meaning of the surface tension parameter $\lambda(=\!1)$  should also be modified  such that $\lambda \to \infty$ for $T\to 0$ for example. Such a change $\lambda\to\infty$ implies that the surface size or more exactly $\langle  S_1\rangle$ changes such that $\langle  S_1\rangle \to 0$ for  $T\to 0$ because $\lambda \langle  S_1\rangle$ remains constant due to the scale invariance of $Z$ described below. Note also that the simulations in this paper are performed  at relatively small range of finite $\kappa$, and hence the surface size remains almost constant.

%----------------------------------------------------------
\subsection{Intrinsic curvature model\label{Int}}
%----------------------------------------------------------
%%%%%%%%%%%%%%%%%%%%%%%%%%%%%%%%%%%%%%%%%%%%%%%%%%%%%%%%%%%%%%%%%%%%
\begin{figure}[ht]
\begin{center}
\includegraphics[width=9.5cm]{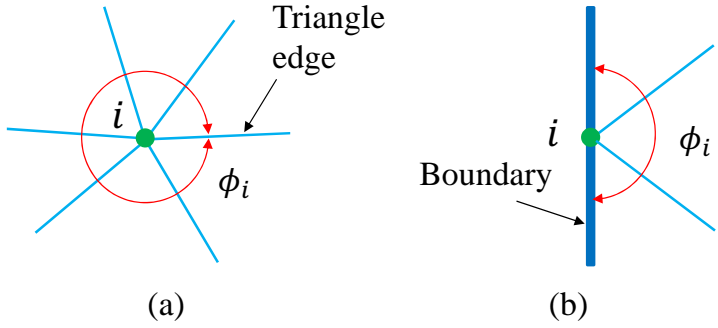}
 \caption{Illustration of the definition of $\phi_i$ of $S_3$ in Eq. (\ref{Hamiltonian-int}) for the intrinsic model. The $\phi_i$ depends on whether the vertex $i$ is (a) internal vertex or (b) boundary vertex.   \label{fig-2} }
\end{center}
\end{figure}
%%%%%%%%%%%%%%%%%%%%%%%%%%%%%%%%%%%%%%%%%%%%%%%%%%%%%%%%%%%%%%%%%%%%
Another model we study is called the {\it intrinsic} curvature model, the Hamiltonian of which is given by \cite{Koibuchi-EPJB2004}
\begin{eqnarray}
\label{Hamiltonian-int} 
&& S({\bf r})=\lambda S_1 + \kappa S_3, \quad(\lambda=1), \nonumber \\
&& S_1=\sum_{ij} \left( {\bf r}_i-{\bf r}_j\right)^2, \quad S_3=\sum_{i=1}^N  \left(\phi_{i}-k\pi \right)^2, \quad ({\rm intrinsic}),\\
&&
\phi_i -k\pi=\left\{ \begin{array}{@{\,}ll}
 \phi_i -2\pi & ({\rm for \; internal\; vertex}\; i) \nonumber \\
 \phi_i -\pi  & ({\rm for \; boundary\; vertex}\; i)
        \end{array} 
         \right.. 
\end{eqnarray} 
The bond potential $S_1$ is identical to that in Eq. (\ref{Hamiltonian-can}), and the intrinsic curvature energy $S_3$ is different from $S_2$ in Eq. (\ref{Hamiltonian-can}). The sum $\sum_i$ in $S_3$ is over all vertices $i$, and $\phi_i$ denotes the sum of the internal angles of triangles meeting at the vertex $i$, and $\phi_i \!-\!k\pi$ depends on whether the vertex $i$ is internal vertex or boundary vertex (Figs. \ref{fig-2}(a),(b)).
 We should note that $2\pi\!-\!\phi_i$ is called "deficit angle" of the internal vertex $i$ \cite{Koibuchi-EPJB2004,Koibuchi-NPB2010}. For the coefficient of $S_3$, we use the same symbol $\kappa$ because the role of this $\kappa$ is almost identical to that in Eq. (\ref{Hamiltonian-can}) for the canonical model. For this reason, this $\kappa$ in Eq. (\ref{Hamiltonian-int}) can also be called bending rigidity. The partition function $Z$ for this intrinsic curvature model is exactly the same as that for the canonical model.

Note that the intrinsic energy $S_3$ in Eq. (\ref{Hamiltonian-int}) is different from the Gaussian curvature term $S_{\rm GC}$ in \cite{HELFRICH-1973}. $S_3$ makes the surface flat by enforcing every deficit angle zero for sufficiently large $\kappa$, while $S_{\rm GC}$ imposes a constraint only on the total sum of internal angles of triangles on the surface (by Gauss-Bonnet theorem \cite{FDavid-SMMS2004}). 

For sufficiently large bending rigidity $\kappa$, the $S_3$ protects the vertices from out-of-plane deformation and does not impose any constraint on in-plane deformation if the surface is flat at least. Therefore, the role of $S_3$ is expected to be almost the same as that of $S_2$ in Eq. (\ref{Hamiltonian-can}), where only out-of-plane deformation is suppressed by large $\kappa$.  We should note that a resistance to shear deformation is still expected in the intrinsic model as well as in the canonical model in contrast to the fluid models as described in Section \ref{tethered-fluid}.   

%----------------------------------------------------------
\subsection{Surface tension\label{tension}}
%----------------------------------------------------------
The surface tension $\sigma$ is calculated from the so-called scale invariance of the partition function for surfaces with fixed boundary \cite{WHEATER-JP1994}, and $\sigma$ is called {\it frame tension} because it is calculated by the projected area $A_p$ of the boundary frame \cite{Cai-Lub-PNelson-JFrance1994}. Note that this $\sigma$ is different from $\lambda$ in Eqs. (\ref{Hamiltonian-can}) and (\ref{Hamiltonian-int}), because $\sigma$ is a physical observable while  $\lambda$ is a "microscopic" surface tension and simply an input parameter. The scale transformation ${\bf r}\to\alpha{\bf r}$ is simply a variable transformation of the integrations in $Z$, and therefore the scaled partition function is
\begin{eqnarray}
\label{scaled-part} 
Z(\alpha)=\alpha^{3N_{\rm E}+2L_2} \int \left(\prod _{i=1}^{N_{\rm E}} d {\bf r}_i\right) \left(\prod _{i=1}^{2L_2} d {\bf r}_i\right)^\prime  \exp\left(-S\left[\alpha{\bf r}, A_p(\alpha)\right]\right)
\end{eqnarray} 
becomes independent of $\alpha$ \cite{WHEATER-JP1994}. The factors $\alpha^{3N_{\rm E}}(=\!\alpha^{3(N-2L_2)})$ and $\alpha^{2L_2}$ come from the three-dimensional and one-dimensional multiple integrations, respectively. We should note that $S$ is considered as a two-component function such as $S\left[\alpha, A_p(\alpha)\right]$. The first $\alpha$ in $S$ denotes the explicit scale parameter such as $\alpha{\bf r}$, and the second $\alpha$ in $A_p(\alpha)$ is an implicit scale parameter, where  $A_p(\alpha)\!=\!\alpha^{-2}A_p$ is assumed. The reason why we assume  $A_p(\alpha)\!=\!\alpha^{-2}A_p$ for the projected area instead of $A_p(\alpha)\!=\!\alpha^{2}A_p$ in $S$ is because the boundary and hence its projected area $A_p$ remain unchanged under ${\bf r}\to\alpha{\bf r}$. Thus, we have a partial derivative formula $[\partial Z/\partial A_p(\alpha)][\partial A_p(\alpha)/\partial\alpha]\!=\!-2A_p\alpha^{-3}\partial Z_{\rm cyl}(A_p)/\partial A_p$. Applying this formula to $\partial \log Z(\alpha) /\partial \alpha\vert_{\alpha=1}\!=\!0$, we have 
\begin{eqnarray}
\label{scale-inv} 
2\langle S_1\rangle-3N+4L_2=-2A_pZ^{-1}\partial Z(A_p)/\partial A_p.
\end{eqnarray}
To evaluate $\partial Z(A_p)/\partial A_p$ in the right-hand side, we assume the partition function
\begin{eqnarray}
\label{free-energy} 
&&Z(A_p)=\exp \left[-\beta F(A_p)\right],\quad (\beta=1),\nonumber \\
&& F(A_p)=\sigma \int_{A_0}^{A_p} dA=\sigma(A_p-A_0).
\end{eqnarray}
 for a macroscopic surface, which spans the boundary of the projected area $A_p$ and has free energy $F(A_p)$ \cite{WHEATER-JP1994}. Thus, we have $\sigma(N)$ as a function of $N$ such that 
\begin{eqnarray}
\label{surface-tension} 
\sigma=\frac{2\langle S_1\rangle-3N+4L_2}{2A_p}.
\end{eqnarray}
We should note that there exists a finite value of $\sigma$, including $\sigma\!\to\! 0$, in the limit of $N\to\infty$. Indeed, $\sigma$ can also be written as $\sigma\!=\![2\langle S_1\rangle\!-\!3N(1\!-\![4/3]L_2/N)]/{(2A_p)}$, where $L_2/N\!\to\! 0 (N\!\to\! \infty)$ for the surfaces with constant $D/L_2$. The projected area $A_p$ is proportional to $N$; moreover, $S_1(N\!\to\! \infty)$ is also proportional to $N$ because of its definition. 

Note also that $\langle S_1\rangle\!=\!\langle \sum_{ij}\ell_{ij}^2\rangle$ becomes constant on the surfaces without the fixed boundaries. Indeed, we have $2\langle S_1\rangle\!-\!3N\!=\!0$ instead of Eq. (\ref{scale-inv}) in this case, because $L_2\!=\!0$ and no constraint for the projected area $A_p$ for surfaces without the fixed boundaries. Thus we obtain $\langle S_1\rangle\!=\!3N/2$. This indicates that the mean bond length squares is given by  $\langle \ell_{ij}^2\rangle\!=\!3N/(2N_B)$, where $N_B\!=\!\sum_{ij}1$ is the total number of bonds. Since $N/N_B$ is constant and independent of $N$, we understand that the mean distance between two neighboring vertices becomes constant even without the fixed boundaries in both surface models.

Here we show that the physical unit of $\sigma$ in Eq. (\ref{surface-tension}) is given by [N/m]. As mentioned in Section \ref{Cano}, the unit of $\langle S_1\rangle$ in Eq. (\ref{surface-tension}) is [1], because the factors $\beta(=\!1)$ and $\lambda(=\!1)$ are suppressed in the expression of $\langle S_1\rangle$ in Eq. (\ref{surface-tension}). Indeed, the unit of  $\lambda \beta \langle S_1\rangle$ is ${\rm [N/m\cdot k_BT\cdot m^2]\!=\![Nm^{-1}\cdot N^{-1}m^{-1}\cdot m^2]\!=\![1]}$. Moreover, due to  the fact that $\beta(=\!1)$ is suppressed in the Boltzmann factor of Eq. (\ref{free-energy}), the denominator $2A_p$ in Eq. (\ref{surface-tension}) has the factor $\beta(=\!1)$  because of the relation in Eq. (\ref{scale-inv}). This proves that the unit of $\sigma$ is [N/m]. 

%----------------------------------------------------------
\subsection{Monte Carlo Technique\label{Monte}}
%----------------------------------------------------------
The standard Metropolis technique is used to update the variables ${\bf r}\!=\!\{{\bf r}_1,{\bf r}_2,\cdots,{\bf r}_N\}$ \cite{Metropolis-JCP-1953,Landau-PRB1976}. As described in the previous subsections, the vertices on the boundary are allowed to move on the boundary circle. This is in contrast to the models in Ref. \cite{Koibuchi-PLA2016}, where the boundary vertices are completely fixed. The reason the boundary points are allowed to move on the circle in the models of this paper is that the fixed boundary condition in Ref. \cite{Koibuchi-PLA2016} is expected to be too strong for surfaces to undergo a continuous transition between the smooth and crumpled (or wrinkled) phases. 

Let ${\bf r}_i^\prime\!=\!{\bf r}_i\!+\!\delta {\bf r}_i$ be a new vertex position with a random three-dimensional vector $\delta {\bf r}_i$ inside a sphere of radius $R_0$. Then the new position ${\bf r}_i^\prime$ is accepted with the probability ${\rm Min}[1,\exp (-\delta S)]$, where $\delta S\!=\!S({\rm new})\!-\!S({\rm old})$ is the change in total energy. The acceptance rate is controlled by $R_0$.

%----------------------------------------------------------
\section{Results\label{results}}
%----------------------------------------------------------
%----------------------------------------------------------
\subsection{Snapshots\label{snapshots}}
%----------------------------------------------------------
%%%%%%%%%%%%%%%%%%%%%%%%%%%%%%%%%%%%%%%%%%%%%%%%%%%%%%%%%%%%%%%%%%%%
\begin{figure}[ht]
\begin{center}
\includegraphics[width=13.5cm]{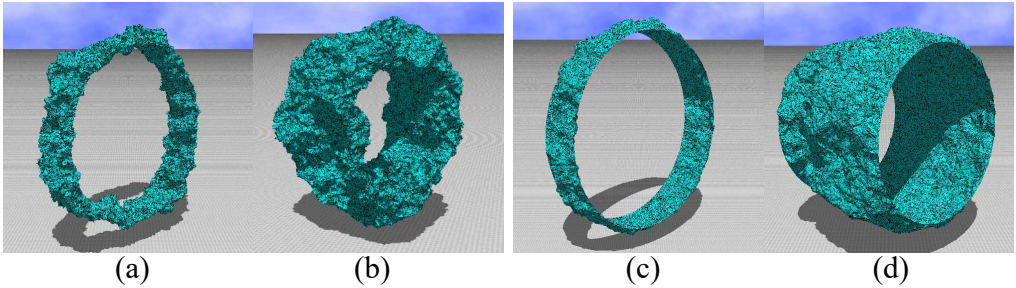}
 \caption{Snapshots obtained by (a),(b) the canonical model and (c),(d) the intrinsic model at the transition points. The sizes are (a),(c) $D/L_1\!=\!4$, $N\!=\!60604$, and (b),(d) $D/L_1\!=\!1$, $N\!=\!59823$. }
 \label{fig-3}
\end{center}
\end{figure}
%%%%%%%%%%%%%%%%%%%%%%%%%%%%%%%%%%%%%%%%%%%%%%%%%%%%%%%%%%%%%%%%%%%%
First, we show lattice snapshots of the canonical model in Figs. \ref{fig-3}(a),\ref{fig-3}(b) and the intrinsic model in Figs. \ref{fig-3}(c) and \ref{fig-3}(d). The lattice size is (a),(c) $D/L_1\!=\!4$, $N\!=\!59823$ and (b),(d) $D/L_1\!=\!1$, $N\!=\!60604$. These snapshots are obtained at the continuous transposition points, where the variance $C_{S_2}$ or $C_{S_3}$, which will be calculated in the next subsection, has the peak. The bending rigidity is given by (a) $\kappa\!=\!0.77$, (b) $\kappa\!=\!0.7608$, (c) $\kappa\!=\!9.3$, and (d) $\kappa\!=\!9.768$. The boundary shape of the canonical model is relatively unclear compared with the intrinsic model. We understand from the snapshots that the intrinsic curvature $S_3$ more strongly suppresses the out-of-plane fluctuations than the extrinsic curvature $S_2$ at the transition point. 

The diameter of the cylinders does not collapses and is almost comparable to the diameter $D(=\!L_2/\pi)$ of the boundary. This is almost clear on the surface of $D/L_1\!=\!4$ (Figs. \ref{fig-3}(a) and \ref{fig-3}(c)), because the distance $L_1$ between the boundaries is relatively smaller than $D$. In the case of $D/L_1\!=\!1$ surfaces (Figs. \ref{fig-3}(a) and \ref{fig-3}(c)), the diameter size still remains around $D$ although the fluctuation of it is relatively large compared to that of the $D/L_1\!=\!4$ surfaces. 

%----------------------------------------------------------
\subsection{Continuous transitions\label{cont-transition}}
%----------------------------------------------------------
%%%%%%%%%%%%%%%%%%%%%%%%%%%%%%%%%%%%%%%%%%%%%%%%%%%%%%%%%%%%%%%%%%%%
\begin{figure}[h]
\begin{center}
\includegraphics[width=10.5cm]{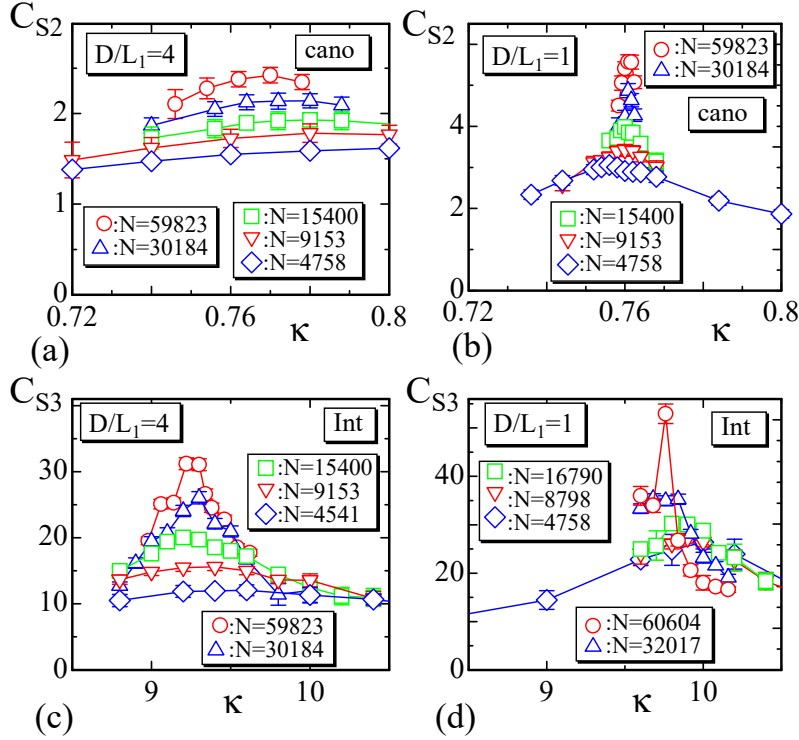}
 \caption{Variance $C_{S_2}$ vs. $\kappa$ of the canonical model for lattices with (a) $D/L_1\!=\!4$ and (b) $D/L_1\!=\!1$, and $C_{S_3}$ vs. $\kappa$ of the intrinsic model for those with (c) $D/L_1\!=\!4$ and (d) $D/L_1\!=\!1$.}
 \label{fig-4}
\end{center}
\end{figure}
%%%%%%%%%%%%%%%%%%%%%%%%%%%%%%%%%%%%%%%%%%%%%%%%%%%%%%%%%%%%%%%%%%%%
We show that the models undergo a continuous transition between the smooth phase and the crumpled or wrinkled phase. Only at the continuous transition point is the surface tension $\sigma$ expected to be dependent on the boundary shape as described above. 
The variances of the curvature energies $S_2$ and $S_3$ are respectively defined by 
\begin{eqnarray}
\label{CS2_CS3} 
C_{S_2}=\frac{1}{N_{\rm E}}\left(\langle S_2^2\rangle - \langle S_2\rangle^2\right),\quad 
C_{S_3}=\frac{1}{N_{\rm E}}\left(\langle S_3^2\rangle - \langle S_3\rangle^2\right),
\end{eqnarray}
where $N_{\rm E}\!=\!N-2L_2$. At the continuous transition point, these quantities have a peak, which is expected to grow with increasing $N$. In Fig. \ref{fig-4}(a), the peaks in $C_{S_2}$ are not so sharp compared to those in Fig. \ref{fig-4}(b). This is because the boundary for the surface with $D/L_1\!=\!4$ strongly suppresses the surface fluctuations in the canonical model. Nevertheless, a weak continuous transition is expected. In the case of intrinsic curvature model, the peaks in $C_{S_3}$ are more clear on the surfaces with $D/L_1\!=\!4$ and $D/L_1\!=\!1$  (Figs. \ref{fig-4}(c) and \ref{fig-4}(d)). We see from the figures that both the surfaces with $D/L_1\!=\!4$ and $D/L_1\!=\!1$ undergo continuous transitions. 

%%%%%%%%%%%%%%%%%%%%%%%%%%%%%%%%%%%%%%%%%%%%%%%%%%%%%%%%%%%%%%%%%%%%
\begin{figure}[h]
\begin{center}
\includegraphics[width=10.5cm]{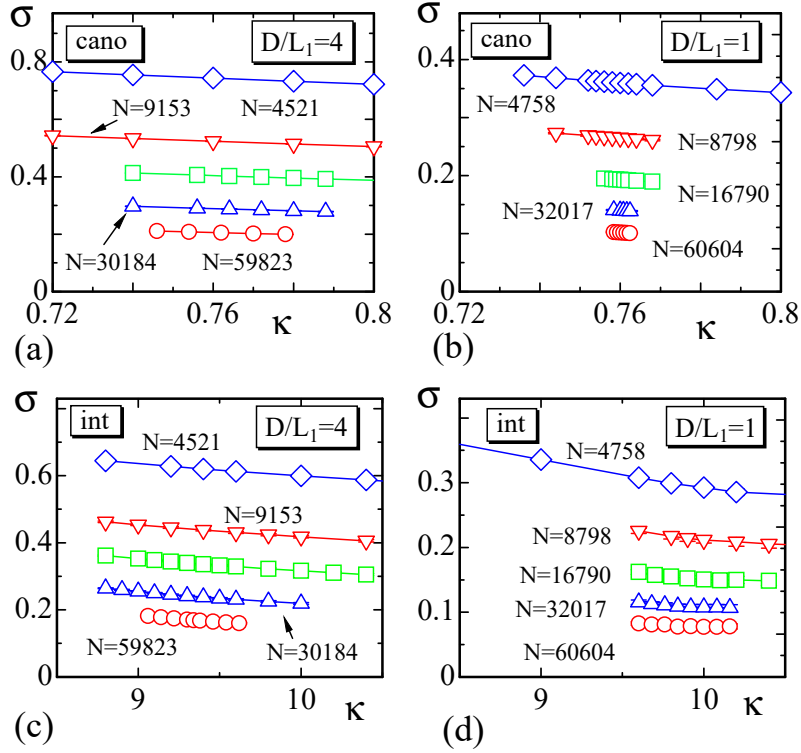}
 \caption{Frame tension $\sigma$ vs. $\kappa$ of the canonical model with (a) $D/L_1\!=\!4$ and (b) $D/L_1\!=\!1$, and $\sigma$ vs. $\kappa$ of the intrinsic curvature model with (c) $D/L_1\!=\!4$ and (d) $D/L_1\!=\!1$. }
 \label{fig-5}
\end{center}
\end{figure}
%%%%%%%%%%%%%%%%%%%%%%%%%%%%%%%%%%%%%%%%%%%%%%%%%%%%%%%%%%%%%%%%%%%%
The frame tension $\sigma$ is plotted against $\kappa$ for the canonical (Figs. \ref{fig-5}(a),(b)) and intrinsic (Figs. \ref{fig-5}(c),(d)) models. The variation of $\sigma$ is smooth for the four different model combinations and $D/L_1$.  We find from these figures that $\sigma$ is decreasing with increasing $N$ as mentioned in the end of Section \ref{tension}. The problem is whether $\sigma$ depends on $D/L_1$, which characterizes the boundary shape, at the transition point in the limit of $N\!\to\!\infty$. This will be shown in the following subsection.

%----------------------------------------------------------
\subsection{Frame tension at the continuous transition point\label{frame-tension}}
%----------------------------------------------------------
%%%%%%%%%%%%%%%%%%%%%%%%%%%%%%%%%%%%%%%%%%%%%%%%%%%%%%%%%%%%%%%%%%%%
\begin{figure}[h]
\begin{center}
\includegraphics[width=10.5cm]{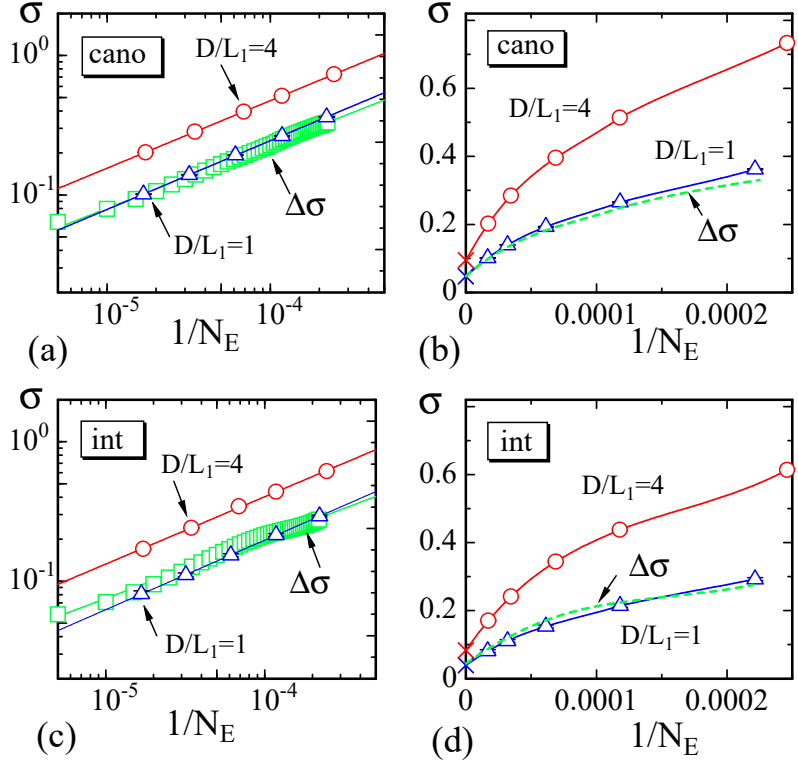}
 \caption{(a), (b) $\sigma$ (${\it \Delta}\sigma$) vs. $1/N_{\rm E}$ obtained at the transition points of the canonical model in the (a) log-log and (b) linear scales. The curves in (b) are drawn by extrapolations \cite{Mathematica}. (c), (d)  $\sigma$  (${\it \Delta}\sigma$) vs. $1/N_{\rm E}$  for the intrinsic model. The crosses ($\times $) plotted at $1/N_{\rm E}\!\to\!0$ in (b) and (d) are obtained by the extrapolations; as a result, the difference ${\it \Delta}\sigma$ (dashed lines in (b), (d)) is found to be nonzero positive in the limit of $1/N_{\rm E}\!\to\!0$.}
 \label{fig-6}
\end{center}
\end{figure}
%%%%%%%%%%%%%%%%%%%%%%%%%%%%%%%%%%%%%%%%%%%%%%%%%%%%%%%%%%%%%%%%%%%%
To see the dependence of $\sigma$ on  $D/L_1$, we plot $\sigma$ of the canonical model in the log-log scale against $1/N_{\rm E}$, as shown in Fig. \ref{fig-6}(a). We also plot the difference ${\it \Delta}\sigma$ defined by 
\begin{eqnarray}
\label{diff-sigma}
{\it \Delta}\sigma=\sigma(D/L_1\!=\!4)-\sigma(D/L_1\!=\!1)).
\end{eqnarray}
Note that $\sigma(D/L_1\!=\!4)$ and $\sigma(D/L_1\!=\!1)$ include the interpolated/extrapolated points shown in Fig. \ref{fig-6}(b). This is the reason why ${\it \Delta}\sigma$ in Fig. \ref{fig-6}(a) has a lot of data points.
 From the slopes of the lines in Fig. \ref{fig-6}(a), $\sigma$ appears to be $\sigma\!\to\!0$  in the limit of $1/N_{\rm E}\!\to\!0$ on both $D/L_1\!=\!4$ and $D/L_1\!=\!1$ surfaces, and ${\it \Delta}\sigma$ also appears to be ${\it \Delta}\sigma(N_{\rm E}\!\to\!\infty)\!\to\!0$. 

However, it is still unclear whether $\sigma$ depends on the boundary shape or not, because  ${\it \Delta}\sigma$ looks out-of-scaling, though $\sigma(D/L_1\!=\!4)$ and $\sigma(D/L_1\!=\!1)$ are parallel. For this reason, we plot $\sigma$ in the linear scale, as shown in Fig. \ref{fig-6}(b). The $\sigma (N_{\rm E}\!\to\!\infty)$ ($\times $) in the limit of $1/N_{\rm E}\!\to\!0$ are obtained by an extrapolation technique \cite{Mathematica}. The results show that both  $\sigma (N_{\rm E}\!\to\!\infty)$ are non-zero and different. In the case of the intrinsic model, relatively the same results occur as with canonical model. The dashed line in (b) denotes ${\it \Delta}\sigma$ defined by Eq. (\ref{diff-sigma}). 

For the data of intrinsic model, the same analyses as those for the canonical model are performed, and the results are plotted in Figs. \ref{fig-6}(c) and \ref{fig-6}(d). Thus, we find that ${\it \Delta}\sigma$ is nonzero positive in the limit of $1/N_{\rm E}\!\to\!0$ in the intrinsic model. The result that $\sigma (N_{\rm E}\!\to\!\infty)$ remains finite indicates that the scaling property shown in (a) and (c) is not exactly satisfied. Moreover, the fact that the difference ${\it \Delta}\sigma(N_{\rm E}\!\to\!\infty)$ is of the order of $\sigma(N_{\rm E}\!\to\!\infty)$ implies that $\sigma(N_{\rm E}\!\to\!\infty)$ remains finite if ${\it \Delta}\sigma(N_{\rm E}\!\to\!\infty)$ does. To support this implication, it is interesting to see whether  ${\it \Delta}\sigma(N_{\rm E}\!\to\!\infty)$ is finite at non transition points. This expectation will be checked below using the  data reported in Ref. \cite{Koibuchi-PLA2016}.

%%%%%%%%%%%%%%%%%%%%%%%%%%%%%%%%%%%%%%%%%%%%%%%%%%%%%%%%%%%%%%%%%%%%
\begin{figure}[h]
\begin{center}
\includegraphics[width=10.5cm]{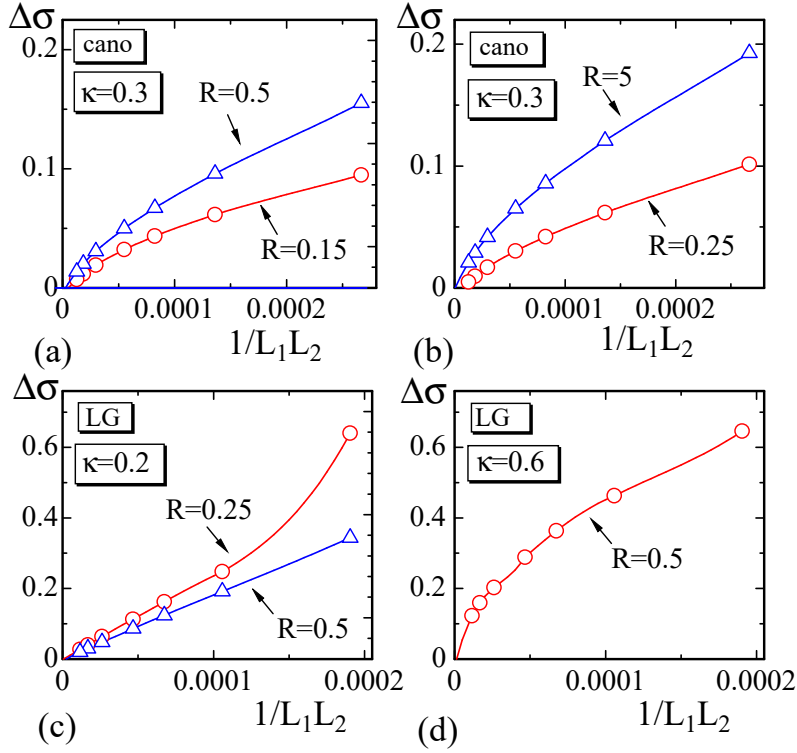}
 \caption{${\it \Delta}\sigma$ vs. $1/L_1L_2$ of (a), (b) the canonical model, and (c), (d) the Landau-Ginzburg model, reported in Ref. \cite{Koibuchi-PLA2016}. The curves are drawn by extrapolations \cite{Mathematica}. ${\it \Delta}\sigma\!\to\!0 \;(1/L_1L_2\!\to\! 0)$ is confirmed in all data plotted. }
 \label{fig-7}
\end{center}
\end{figure}
%%%%%%%%%%%%%%%%%%%%%%%%%%%%%%%%%%%%%%%%%%%%%%%%%%%%%%%%%%%%%%%%%%%%
Thus, the next task is to plot the data ${\it \Delta}\sigma (=\!|\sigma_I\!-\!\sigma_{II}|)$ of Ref. \cite{Koibuchi-PLA2016} against $1/L_1L_2\!\to\!0$. In Ref \cite{Koibuchi-PLA2016}, $1/L_1L_2$ is used for the log-log plots instead of $1/N_{\rm E}$, which is used in Figs. \ref{fig-7}(b) and \ref{fig-7}(d). In the case of Ref. \cite{Koibuchi-PLA2016}, $\sigma_I$ and $\sigma_{II}$ are obtained on two different boundary shape lattices (see Ref. \cite{Koibuchi-PLA2016} in more detail). In Ref. \cite{Koibuchi-PLA2016}, all data were obtained at the non-transition region; therefore, the value $\sigma$ itself is very large in contrast to the data at the transition point in this paper. For this reason, the difference ${\it \Delta}\sigma$ is plotted in Fig. \ref{fig-7} (and also in Ref. \cite{Koibuchi-PLA2016}). The symbols "cano" and "LG" on the figures denote the canonical and Landau-Ginzburg models, of which the results were presented in Ref. \cite{Koibuchi-PLA2016}. $R\!=\!A_p/L_1L_2$ is called expansion ratio in Ref. \cite{Koibuchi-PLA2016}. We confirm from Figs. \ref{fig-7}(a)-(d) that ${\it \Delta}\sigma\!\to\!0$ in the limit of $1/L_1L_2\!\to\!0$. The fact that ${\it \Delta}\sigma(L_1L_2\!\to\!\infty)\!\to\!0$ indicates that the scaling of ${\it \Delta}\sigma$ with respect to $L_1L_2$  is exact in contrast to the data at the transition points shown in Figs. \ref{fig-6}(a), (c). 
This implies that the frame tension is independent of the boundary shape at the non-transition region. This conclusion, already reported in Ref. \cite{Koibuchi-PLA2016}, is reconfirmed by the technique introduced for analyzing the results obtained in this paper.

%----------------------------------------------------------
\section{Summary and Conclusion\label{conclusion}}
%----------------------------------------------------------
We used Monte Carlo simulations to study an expectation that the surface (or frame) tension of membranes depends on the boundary shape at the continuous transition point between the smooth and wrinkled phases. This is expected from an analogy of the phase transition theoretical viewpoint with spin models; however, this is a nontrivial problem because surface normal vectors are different from the spins in spin models. To confirm this expectation numerically, we use two different models defined on triangulated surfaces. We find in both models that the frame tension $\sigma$ with one boundary shape is different from $\sigma$ calculated on surfaces with the other boundary shape at the continuous transition point. This implies the possibility that mechanical properties of isotropic materials depend on their shape.

{\bf Acknowledgment}
This work is supported in part by JSPS KAKENHI, Grant No. 17K05149.

\end{document}